# Research on Event-Related Desynchronization of Motor Imagery and Movement Based on Localized EEG Cortical Sources

Yuqing Wang

(Chongqing University-University of Cincinnati Joint Co-op Institute，Chongqing，400044)

**Abstract:** This study investigates event-related desynchronization (ERD) phenomena during motor imagery and actual movement. Using sLORETA software, we analyzed the cortical current source density distributions in Mu and Beta frequency bands for 33 subjects during rest, motor imagery, and actual movement conditions. The results were normalized for analysis. Using sLORETA's statistical tools, paired t-tests were conducted to compare the normalized current source density results between rest and motor imagery, rest and actual movement, and motor imagery and actual movement conditions in both frequency bands. The findings revealed: In both Mu and Beta frequency bands, during motor imagery, significant ERD ($P<0.01$) was observed in the salience network, supplementary motor area, primary motor area, premotor cortex, primary somatosensory cortex, and parietofrontal mirror neuron system. During actual movement, significant ERD ($P<0.05$) was observed in the primary somatosensory cortex, primary motor area, and parietofrontal mirror neuron system in both frequency bands. Comparing motor imagery to actual movement, the current source density in the primary somatosensory cortex and parietofrontal mirror neuron system was higher during motor imagery, though this difference was not statistically significant ($P>0.05$). This paper analyzes the factors contributing to these statistical results and proposes preliminary solutions.

# 1 Introduction

Motor imagery refers to the process where subjects imagine performing a specific action without generating any physical movement[1]. Related studies have shown that the brain regions activated during motor imagery are similar to those activated during motor execution, both in terms of location and intensity. These regions include the primary somatosensory cortex, premotor cortex, supplementary motor area (SMA), primary motor area, callosal convolution, parietal cortex, thalamus, and cerebellum[2]. The main difference lies in the functional connectivity patterns between these regions during different tasks.

When a specific region of the cerebral cortex receives movement commands or imagines movement, the area becomes activated, leading to increased metabolism and information processing. This results in a decrease in the amplitude of EEG signals in specific frequency bands, a physiological phenomenon known as event-related desynchronization (ERD). Conversely, when an activity does not significantly activate related cortical areas at a particular time, the EEG signals in specific frequency bands show increased amplitude, known as event-related synchronization (ERS). Numerous studies have shown that during actual movement or motor imagery, the Mu (8-12Hz) and Beta (18-25Hz) rhythms in the sensorimotor cortex and parietal regions exhibit power attenuation (ERD)[3]. Typically, brain-computer interface (BCI) systems' research on motor imagery signals is based on the event-related desynchronization phenomenon of these rhythmic components.

Current EEG technology can directly measure neural electrical signals from the scalp with high temporal resolution. However, due to the brain's volume conduction effect, scalp EEG signals have low spatial resolution, and signals from different electrodes are coupled, making it challenging to locate activated brain regions and determine their activation levels[4]. Source localization algorithms can reconstruct brain current density distribution by inversely estimating the location, direction, and intensity of neural signal sources based on scalp potential values, thereby analyzing the spatiotemporal characteristics of brain activity.

To address the problems of low spatial resolution and strong coupling between adjacent channels in scalp EEG signals, this paper uses source analysis methods to study the differences in deep neural current source density distribution among thirty-three subjects during rest, motor imagery, and actual movement. This approach enables investigation of event-related desynchronization phenomena associated with motor imagery and actual movement.

# 2 Experimental Data and Methods

## 2.1 Experimental Data

The experimental data was provided by the New York State Department of Health and State University of New York[5].

The experimental task involved 33 subjects (19 males, 14 females, mean age 39.6 years) completing "motor tasks" and "imagery tasks" according to visual cues. All subjects were seated in an adjustable armchair facing a computer display. Subjects were instructed to avoid blinking and large body movements during the experiment.

The complete experimental process consisted of 6 rounds. In 3 rounds, subjects were required to imagine left- or right-hand movement according to visual cues. In the other 3 rounds, subjects were required to open and close their left or right hand according to visual cues. Each round lasted two minutes, with a one-minute rest period between rounds. Each round included 15 tasks lasting 4 seconds each, during which a vertical line appeared on either the left or right edge of the computer display. Subjects were instructed to imagine movement of the corresponding hand or perform actual opening and closing of that hand. Between each task was a 4-second transition period where the display was blank, and subjects were required to relax their entire body without performing imagery or motor tasks.

EEG signal acquisition used the international standard 10-10 system[6], with the reference electrode placed on the right mastoid. The AD sampling rate was 128 Hz, with a bandpass filter of 1-60 Hz.

### 2.2 Source Analysis Method

Standardized Low Resolution Brain Electromagnetic Tomography (sLORETA) can utilize EEG data recorded from the scalp to infer the neural electrical activity sources within the cerebral cortex through inverse solution calculations. To obtain deep signal sources and analyze the changes in current density across relevant brain regions during rest, motor imagery, and actual movement, this paper employs the sLORETA source localization algorithm[7] to inversely estimate the location, orientation, and intensity information of neural signal sources, thereby reconstructing the brain current density distribution.

The three-dimensional linear decomposition current density calculated by the sLORETA algorithm is consistent with the Talairach probabilistic brain atlas model and displays three-dimensional images through MNI152 coordinates at 5 mm resolution. This algorithm assumes that neural sources are distributed in the cortical layer, dividing it into 6239 dense cubic grid points, with each grid point containing a dipole fixed in position but with unknown dipole moment. The measured values at external field points are the superposition of the effects of equivalent dipoles at each grid point within the cortical neural source space. Since the positions of individual small dipoles are known, there exists a linear relationship between neural sources and scalp electrical field distribution. The lead field is used to relate the three dipole moments of any dipole at each time point to the electrical field values at each electrode:

$$\Phi = KJ \qquad (1)$$

Here, is a column vector composed of scalp potential values at E measurement electrodes; is a three-dimensional column vector composed of current densities at known grid points in the brain; at the i-th voxel, it contains three unknown dipole moments; K is the lead field matrix, which represents the relationship between scalp field values and internal neural sources.

This equation is underdetermined and has no unique solution. A cost equation is introduced:

Here, $\Phi = [\Phi_1, \Phi_2 \ldots \Phi_{N_E}]^T$ is a $N_E$ column vector composed of scalp potential values at E measurement electrodes; $J = (J_1^T, J_2^T, J_3^T \ldots J_{N_V}^T)^T$ is $3N_V$ column vector, composed of current densities at $N_V$ known grid points in the brain; At $l-th$ voxel, $J_l^T = (J_l^x, J_l^y, J_l^z)$ ($J_l$ is a three $-$ dimensional column vector, l = 1,2 \ldots, $N_V$), $J_l^T$ contains three unknown dipole moments; K is the $N_E \times 3N_V$ lead field matrix, which represents the relationship between scalp field values and internal neural sources.。

This equation is underdetermined and has no unique solution. A cost equation is introduced:

$$F = \|\Phi - KJ\|^2 + \alpha\|J\|^2 \tag{2}$$

where $\alpha \geq 0$ is the regularization parameter。When solving the EEG inverse problem, with K, $\Phi$ and $\alpha$ known, minimizing the cost equation yields the solution for J:

$$\begin{aligned} J' &= T\,\Phi \\ T &= K^T[KK^T + \alpha H]^+ \\ H &= I - 11^T/1^T1 \end{aligned} \tag{3}$$

Here, $H \in R^{N_E \times N_E}$, is a symmetric and idempotent matrix;; $I \in R^{N_E \times N_E}$ is an identity matrix; $I \in R^{N_E \times 1}$ is an N $-$ dimensional column vector composed of 1s。For any matrix M, $M^+$ denotes the Moore-Penrose generalized inverse。H is the average reference operator。$\Phi$ represents the result of average reference transformation of the measured scalp potential values, and K represents the lead field matrix after average reference transformation.

J′ is the estimated value of current density。It still needs to be standardized using variance. The actual variance of brain current source density：$S_J \in R^{3N_V \times 3N_V}$ equals the identity matrix, i.e.:

$$S_J = I, \quad I \in R^{3N_V \times 3N_V} \tag{4}$$

Part of the variance in scalp potential comes from measurement noise:

$$S_\Phi^{noise} = \alpha H \tag{5}$$

Since measurement noise and neuronal activity are independent, the variance of scalp potential $S_\Phi \in R^{N_E \times N_E}$ is:

$$S_\Phi = KS_JK^T + S_\Phi^{noise} = KK^T + S_\Phi^{noise} = KK^T + H \tag{6}$$

The variance of the estimated current density $J'$ is derived as:

$$S_{J'} = TS_\Phi T^T = T(KK^T + H)T^T = K^T[KK^T + H]^+ K \qquad (7)$$

Finally, the standardized current density is obtained:

$$J'^T_l \left\{[S_{J'}]_{ll}\right\}^{-1} J'_l \qquad (8)$$

$J_l \in R^{3\times 1}$ is the estimated current density of the i-th voxel obtained from equation (3); $[S_{J'}]_{ll} \in R^{3\times 3}$ is the $l-$th diagonal block matrix in equation (7). For a time series of duration $t = 1, \ldots N_T$, equation (8) can be generalized as:

$$J'^T_{l(\forall t)} \left\{[S_{J'}]_{ll}\right\}^{-1} J'_{l(\forall t)} \qquad (9)$$

To solve for the standardized current source density distribution of an EEG segment at a specific frequency $\omega$, first construct the cross-spectral matrix as shown in equation (10) [8]. Where $[\Phi_\omega]_i \in C^{N_E \times 1}$, $i = 1, \ldots N$ represents: the discrete Fourier transform at frequency $\omega$ of the i-th EEG Epoch.

$$\begin{aligned} &_i = T[\Phi_\omega]_i \\ \frac{1}{a}\sum_{i=1}^{N}\left[J'_\omega\right]_i \left[J'_\omega\right]_i^* &= \frac{1}{a}T\left\{\sum_{i=1}^{N}[\Phi_\omega]_i [\Phi_\omega]_i^*\right\}T^T \\ a &= 2\times \pi \times N \times T \end{aligned} \qquad (10)$$

For example, assuming each EEG Epoch contains 256 discrete time points with a sampling rate of 128Hz, this means each EEG Epoch is 2 seconds long. The superscript '*' represents the conjugate transpose matrix. Equation (10) is equivalent to:

$$\begin{aligned} A_{f_\omega} &= TA_{\phi_\omega} T^T \\ A_{f_\omega} &= \frac{1}{a}\sum_{i=1}^{N}\left[J'\omega\right]_i \left[J'\omega\right]_i^* \\ A_{\phi_\omega} &= \sum_{i=1}^{N}[\Phi_\omega]_i [\Phi_\omega]_i^* \end{aligned} \qquad (11)$$

$A_{J'_\omega}$ and $A_{\phi_\omega}$ respectively represent the cross-spectral matrices of current density and scalp potential difference. They are both Hermitian matrices. At frequency $\omega$, we are concerned with $\text{diag}(A_{J'_\omega}) = \text{diag}(TA_{\phi_\omega}T^T)$. The 'diag' operator acts on a Hermitian matrix, extracting its diagonal elements to construct a real diagonal matrix. Then, using the same principles as equations (3)-(8), we can solve for the standardized current density value of the i-th voxel at frequency $\omega$ as shown in equation (12), where $J'_{(\omega)} \in C^{3N_V \times N_T}$.

$$J'^{T}_{l(\omega)} \left\{ [S_{J'}]_{ll} \right\}^{-1} J'_{l(\omega)} \qquad (12)$$

This study used sLORETA software to analyze the brain current source density distribution in the Mu (8-12Hz) and Beta (18-25Hz) frequency bands during rest, motor imagery, and motor execution in 33 subjects. All source localization results were normalized [9].

### 2.3 Statistical Analysis

Using the statistical tools in sLORETA software, paired t-tests were performed on the normalized current source density results between rest and motor imagery, rest and motor execution, and motor imagery and motor execution under two frequency bands, yielding statistical parameter values for brain current density distribution. The logarithm of the F-ratio mean was used to evaluate the degree of difference in current density. The null hypothesis was rejected when P<0.05, indicating significant differences between the two samples. Statistical P-values for multiple comparisons were corrected using the SnPM program[10]. This was used to determine: whether there were significant differences between brain regions activated during motor imagery versus rest; whether there were significant differences between brain regions activated during actual movement versus rest; and whether there were significant differences between brain regions activated during motor imagery versus actual movement.

### 3 Statistical Results

Subjects showed significant differences in brain current source density in both Mu and Beta frequency bands during rest versus motor imagery, and rest versus motor execution, as shown in Table 1.

Figure 1 shows the statistical probability maps (including 3D and sectional views) reconstructed from corresponding log of F-ratios values with P<0.01, which more intuitively displays the differences in brain current source density distribution between rest and motor imagery in both Mu and Beta frequency bands. Figure 2 shows the statistical probability maps (including 3D and sectional views) reconstructed from corresponding log of F-ratios values with P<0.05, which more intuitively displays the differences in brain current source density distribution between rest and motor imagery in both Mu and Beta frequency bands. The red regions indicate positive log of F-ratios values, meaning that the current density values in these brain regions were significantly higher during rest compared to during motor imagery.

Figure 3 shows the 3D and sectional views of brain regions with differences in current source density between motor imagery and actual movement in both Mu and Beta frequency bands. The red regions indicate positive log of F-ratios values, meaning that the current density values in these brain regions were higher during motor imagery compared to during actual movement; the blue regions indicate negative log of F-ratios values, meaning that the current density values in these brain regions were higher during actual movement compared to during motor imagery.

Table 1 Statistical Comparison of Current Source Density in Mu and Beta Frequency Bands during Rest versus Motor Imagery, and Rest versus Actual Movement

| | X-MNI | Y-MNI | Z-MNI | Brodmann Area | Structure | Log of F-ratios |
|---|---|---|---|---|---|---|
| Statistical Comparison of Current Source Density in Mu Frequency Band between Rest and Motor Imagery (P<0.01) | -5 | -35 | 0 | 32 | Anterior Cingulate Gyrus | 0.331088 |
| | 10 | 40 | -5 | 10 | Medial Frontal Gyrus | 0.299845 |
| | -35 | -30 | 40 | 2 | Inferior Parietal Gyrus | 0.283085 |
| | -35 | -25 | 40 | 3 | Postcentral Gyrus | 0.281699 |
| | -35 | -35 | 40 | 40 | Inferior Parietal Lobule | 0.271396 |
| | 30 | -20 | 45 | 4 | Precentral Gyrus | 0.247778 |
| | -35 | 10 | 35 | 6 | Precentral Gyrus | 0.234849 |
| | -15 | -50 | 40 | 7 | Precuneus | 0.22448 |
| Statistical Comparison of Current Source Density in Beta Frequency Band between Rest and Motor Imagery (P<0.01) | -30 | -20 | 45 | 4 | Precentral Gyrus | 0.246472 |
| | -30 | -25 | 45 | 3 | Postcentral gyrus | 0.246067 |
| | -35 | -40 | 45 | 2 | Postcentral gyrus | 0.222478 |
| | -35 | -15 | 45 | 6 | Precentral Gyrus | 0.21877 |
| | -30 | -10 | 45 | 6 | Middle frontal gyrus | 0.209485 |
| Statistical Comparison of Current Source Density in Mu Frequency Band | -35 | -30 | 40 | 2 | Sub gyral | 0.439453 |
| | -35 | -25 | 40 | 3 | Postcentral Gyrus | 0.423481 |
| | -35 | -35 | 40 | 40 | Inferior Parietal | 0.417926 |

| | | | | | Lobule | |
|---|---|---|---|---|---|---|
| between Rest and Actual Movement (P<0.05) | | | | | | |
| Statistical Comparison of Current Source Density in Beta Frequency Band between Rest and Actual Movement (P<0.05) | -30 | -25 | 45 | 3 | Postcentral Gyrus | 0.413471 |
| | -30 | -20 | 45 | 4 | Precentral Gyrus | 0.401275 |

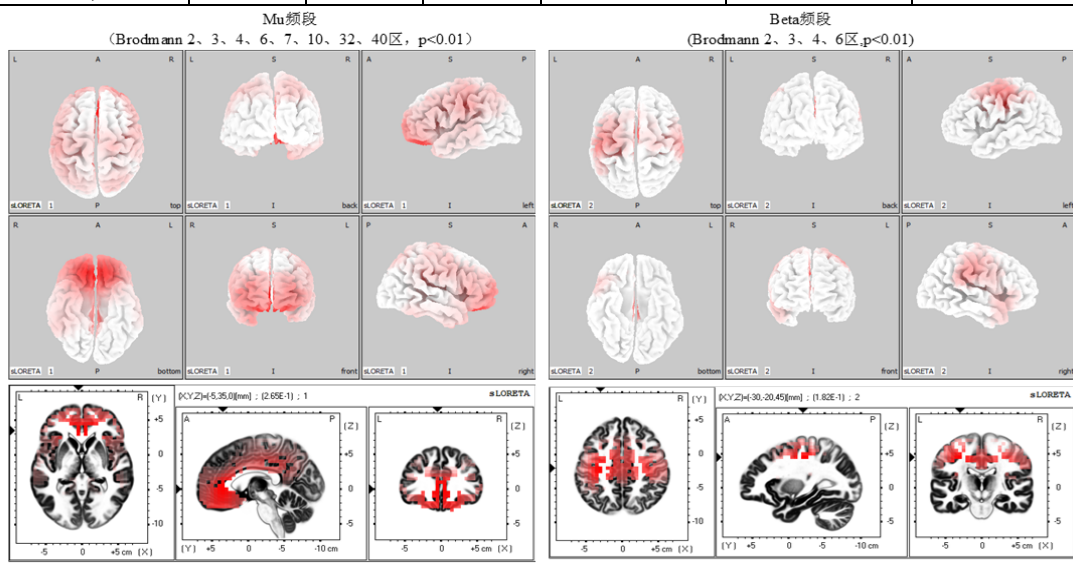

Figure 1 Three-dimensional and sectional views of brain regions showing statistically significant differences in current source density between rest and motor imagery in Mu and Beta frequency bands

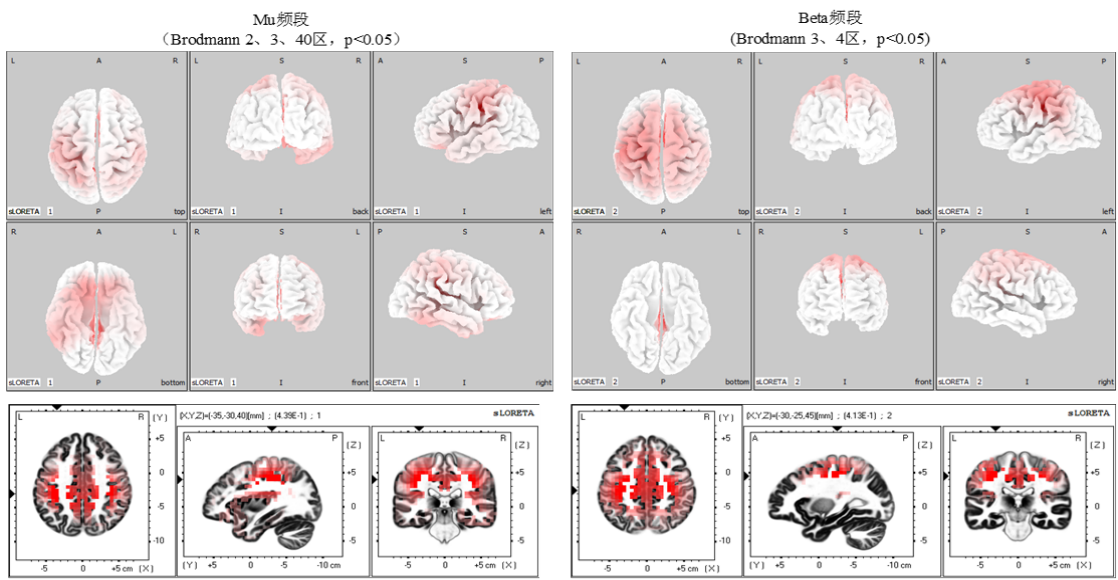

Figure 2 Three-dimensional and sectional views of brain regions showing statistically

significant differences in current source density between rest and actual movement in Mu and Beta frequency bands

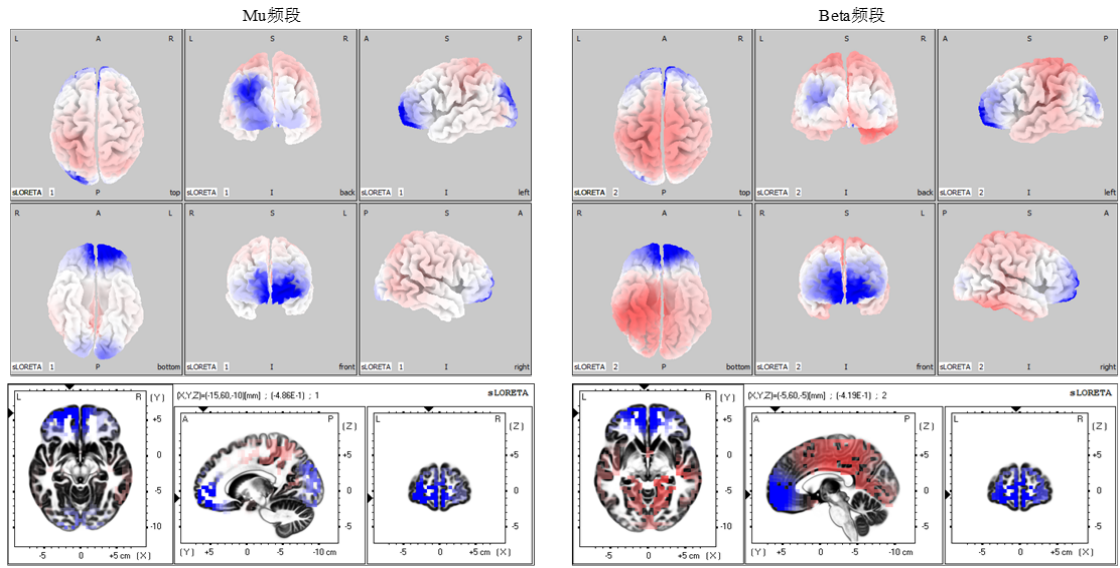

Figure 3 Three-dimensional and sectional views of brain regions showing differences in current source density between motor imagery and actual movement in Mu and Beta frequency bands

## 4　Discussion

In the Mu frequency band, there were significant differences in brain current source density distribution between rest and motor imagery (P<0.01), as shown in Table 1. The brain current source density in Brodmann areas 2, 3, 4, 6, 7, 10, 32, and 40 was significantly lower during motor imagery compared to rest, indicating significant desynchronization in these regions. Brodmann area 32 is part of the salience network, which helps people detect salient external events and facilitates attention, working memory acquisition, and motor preparation [11]. The desynchronization in this region indicates that the salience network was activated during motor imagery. The precuneus is closely related to attention control, and the desynchronization in Brodmann area 7 indicates significant activation of the somatosensory cortex. The precentral and postcentral gyri are closely related to motor control, and the desynchronization in Brodmann areas 2, 3, 4, and 6 indicates significant activation of the primary and secondary sensorimotor cortex, primary motor area, premotor area, and supplementary motor area. Motor imagery involves less visual-motor transformation processing, so the desynchronization in Brodmann area 40 may reflect higher-level cognitive and motor functions[12]. Additionally, this region is part of the parietal-frontal neuron system[13], indicating that the parietal-frontal neuron system was activated during motor imagery.

In the Beta frequency band, there were significant differences in brain current source density distribution between rest and motor imagery (P<0.01), as shown in Table 1. The brain current source density in Brodmann areas 2, 3, 4, and 6 was significantly lower during motor imagery compared to rest, indicating significant desynchronization in these regions. This suggests that during motor imagery, the primary and secondary sensorimotor cortex, supplementary motor area, primary motor area, and premotor area were significantly activated.

In the Mu frequency band, there were significant differences in brain current source density distribution between rest and motor execution (P<0.05), as shown in Table 1. The brain current source density in Brodmann areas 2, 3, and 40 was significantly lower during actual movement compared to rest, indicating significant desynchronization in these regions. This suggests that during actual movement, the primary and secondary sensorimotor cortex and the parietal-frontal mirror neuron system were significantly activated.

In the Beta frequency band, there were significant differences in brain current source density distribution between rest and motor execution (P<0.05), as shown in Table 1. The brain current source density in Brodmann areas 3 and 4 was significantly lower during actual movement compared to rest, indicating significant desynchronization in these regions. This suggests that during actual movement, the primary and secondary sensorimotor cortex and primary motor area were significantly activated.

In both Mu and Beta frequency bands, there were differences in brain current source density distribution between motor imagery and motor execution, but these differences were not significant, as shown in Figure 3 (P>0.05). However, in both Mu and Beta frequency bands, the current source density values were higher in Brodmann areas 2, 3, 7, and 40 during motor imagery. This indicates that event-related desynchronization of Mu rhythm and Beta rhythm in these brain regions was more pronounced during actual movement. Previous research has shown that the activation level during motor execution is significantly higher than during motor imagery[14].

The reasons for non-significant statistical results may be as follows:

1.The solution space of sLORETA is based on a realistic head model derived from MRI images of 152 subjects at the Montreal Neurological Institute. In reality, each subject's actual head structure has subtle differences from this model, so the brain current source density distribution calculated by the sLORETA algorithm may contain errors. These errors could be reduced by using registration with each subject's individual MRI images.

2.The sLORETA algorithm divides the cerebral cortex into 6239 voxels, assuming each voxel represents one neuron. In reality, the number of neurons in the brain is greater than 6239, so the calculated brain current source density distribution has low resolution, which may affect the results of statistical tests.

3.The significance testing in sLORETA software is based on voxel-by-voxel statistical testing. According to cognitive psychology, activated brain regions consist of multiple clusters of voxels with similar activity levels, and differences between individual voxels may cancel each other out.

Preliminary improvement methods could include conducting statistical tests on a cluster-by-cluster basis, a method commonly used in fMRI result statistical testing [15]; or representing each Brodmann area with a central voxel and then conducting statistical tests between these central voxels.

## Reference

[1] Jeannerod M. Mental imagery in the motor context. *Neuropsychologia*. 1995;33(11):1419-1432. doi:10.1016/0028-3932(95)00073-c

[2] Zhizeng LUO, Xianju LU, Ying ZHOU. EEG Feature Extraction Based on Brain Function Network and Sample Entropy[J]. Journal of Electronics & Information Technology, 2021, 43(2): 412-418. doi: 10.11999/JEIT191015

[3] Avanzini, P., Fabbri-Destro, M., Dalla Volta, R., Daprati, E., Rizzolatti, G., & Cantalupo, G. (2012). The dynamics of sensorimotor cortical oscillations during the observation of hand movements: an EEG study. *PloS one*, *7*(5), e37534.

[4] Schoffelen, J. M., & Gross, J. (2009). Source connectivity analysis with MEG and EEG. *Human brain mapping*, *30*(6), 1857-1865.

[5] Nomenclature, S. E. P. (1991). American electroencephalographic society guidelines for. *Journal of clinical Neurophysiology*, *8*(2), 200-202.

[6] Nomenclature, S. E. P. (1991). American electroencephalographic society guidelines for. *Journal of clinical Neurophysiology*, *8*(2), 200-202.

[7] Pascual-Marqui, R. D. (2002). Standardized low-resolution brain electromagnetic tomography (sLORETA): technical details. *Methods Find Exp Clin Pharmacol*, *24*(Suppl D), 5-12.

[8] Pascual-Marqui, R. D. (2002). Notes on some physics, mathematics, and statistics for time and frequency domain LORETA.

[9] Babiloni, C., Cassetta, E., Binetti, G., Tombini, M., Del Percio, C., Ferreri, F., ... & Rossini, P. M. (2007). Resting EEG sources correlate with attentional span in mild cognitive impairment and Alzheimer's disease. *European Journal of Neuroscience*, *25*(12), 3742-3757.

[10] Nichols, T. E., & Holmes, A. P. (2002). Nonparametric permutation tests for functional neuroimaging: a primer with examples. *Human brain mapping*, *15*(1), 1-25.

[11] Yin, S., Liu, Y., & Ding, M. (2016). Amplitude of sensorimotor mu rhythm is correlated with BOLD from multiple brain regions: a simultaneous EEG-fMRI study. *Frontiers in Human Neuroscience*, *10*, 364.

[12] DENG Xin, XIAO Lifeng, YANG Pengfei, et al. Development of a robot arm control system using motor imagery electroencephalography and electrooculography [J]. CAAI Transactions on Intelligent Systems, 2022, 17(6): 1163-1172. doi: 10.11992/tis.202107042

[13] CUI Yao; CONG Fang; LIU Lin. Basic Theory of Mirror Neuron System and Its Meanings in Motor Rehabilitation (review)[J]. 《Chinese Journal of Rehabilitation Theory and Practice》, 2012, 18(3): 239-243.

[14] Hu Y, Liu Y, Cheng C, Geng C, Dai B, Peng B, Zhu J, Dai Y. [Multi-task motor imagery electroencephalogram classification based on adaptive time-frequency common spatial pattern combined with convolutional neural network]. Sheng Wu Yi Xue Gong Cheng Xue Za Zhi. 2022 Dec 25;39(6):1065-1073. Chinese. doi: 10.7507/1001-5515.202206052. PMID: 36575074; PMCID: PMC9927174.

[15] Hanakawa, T., Immisch, I., Toma, K., Dimyan, M. A., Van Gelderen, P., & Hallett, M. (2003). Functional properties of brain areas associated with motor execution and imagery. *Journal of neurophysiology*, *89*(2), 989-1002.